\documentclass{llncs}
 
\usepackage{amsmath,amssymb}
\usepackage{QED}
\usepackage[all]{xy}
\usepackage{prooftree}

\usepackage{pstricks,pst-node}
 \psset{rowsep=.1cm,colsep=.1cm}
 \psset{arrowsize=3.5pt 3.8,arrowlength=0.7}
 \psset{labelsep=.1cm,nodesep=2pt}
\newtheorem{defn}{Definition}

\newtheorem{lem}[defn]{Lemma}

\newtheorem{nota}{Notation}

\author{St\'ephane Le Roux\thanks{http://perso.ens-lyon.fr/stephane.le.roux/. Now working at INRIA-Microsoft Research. I thank Pierre Lescanne for his comments on the draft of this paper.}}
 
\institute{\'Ecole normale sup\'erieure de Lyon, Universit\'e de Lyon, LIP, CNRS, INRIA, UCBL}

\title{Discrete Non Neterminism and Nash Equilibria for Strategy-Based Games}

\def\_{\kern.08em\vbox{\hrule width.35em height.6pt}\kern.08em}

\newcommand{\coqdocindent}[1]{\noindent\kern#1}

\newcommand{\triple}[3] {\langle {#1}, {#2}, {#3}\rangle}
\newcommand{\quadruple}[4] {\langle {#1}, {#2}, {#3}, {#4}\rangle}
\newcommand{\eqdef}{\stackrel{\Delta}{=}}

\newcommand{\powerset}[1] {{\mathcal P}({#1})}

\newcommand{\Agent}{\mathcal{A}}

\newcommand{\agent}{\mathrm{a}}

\newcommand{\V}{V}
\newcommand{\Oc}{Oc}

\newcommand{\node}{n}

\newcommand{\aPrefFct}[2] {\lhd_{#2}^{#1}}
\newcommand{\aPref}[4]     {#3 \mathop{\aPrefFct{#1}{#2}} #4}

\newcommand{\aConvFct}[2]{\ensuremath{\mathrel{\xymatrix@C3pt{\ar@{>-}^{#1}_{#2}[r]&}}}\!\!}

\newcommand{\aConvEqFct}[2]{\ensuremath{\mathrel{\xymatrix@C3pt{\ar@{>-<}^{#1}_{#2}[r]&}}}}
\newcommand{\aConvEq}[4]        {{{#3}} \aConvEqFct{{#1}}{{#2}} {{#4}}}
\newcommand{\lAgentConvFct}[2]{\ensuremath{\mathop{\xymatrix@C3pt{\ar@{>-}^{#1}_{#2}[r]&}}}}

\newcommand{\EqFctName}[2]      {\mathrm{Eq}^{#1}_{#2}}
\newcommand{\EqFct}[3]          {\EqFctName{{#1}}{{#2}}({#3})}

\DeclareMathSymbol{\leggedrightarrow}{\mathord}{AMSa}{"4B}
\newcommand{\simS}{\mathrm{s}}

\DeclareMathSymbol{\leggedrightarrow}{\mathord}{AMSa}{"4B}

\begin{document}

\maketitle

\begin{abstract}
Several notions of game enjoy a Nash-like notion of equilibrium without guarantee of existence. There are different ways of weakening a definition of Nash-like equilibrium in order to guarantee the existence of a weakened equilibrium. Nash's approach to the problem for strategic games is probabilistic, \textit{i.e.} continuous, and static. CP and BR approaches for CP and BR games are discrete and dynamic. This paper proposes an approach that lies between those two different approaches: a discrete and static approach. multi strategic games are introduced as a formalism that is able to express both sequential and simultaneous decision-making, which promises a good modelling power. multi strategic games are a generalisation of strategic games and sequential graph games that still enjoys a Cartesian product structure, \textit{i.e.} where agent actually choose their strategies. A pre-fixed point result allows guaranteeing existence of discrete and non deterministic equilibria. On the one hand, these equilibria can be computed with polynomial (low) complexity. On the other hand, they are effective in terms of recommendation, as shown by a numerical example.
\end{abstract}


\section{Introduction}

Not all strategic games have a (pure) Nash equilibrium. On the one hand, Nash's probabilistic approach copes with this existence problem with an ad hoc solution: Nash's solution is dedicated to a setting with real-valued payoff functions. On the other hand, CP and BR games propose an abstract and general approach that is applicable to many types of game. Both approaches generalise the notion of Nash equilibrium and guarantee the existence of a weakened Nash equilibrium. There are two main differences between the two approaches though. First, Nash's approach considers finite objects and yields continuous objects, whereas the CP and BR approach preserves finiteness. Second, Nash's approach is static, whereas the CP and BR approach is dynamic: Nash's approach is static because a probabilistic Nash equilibrium can be interpreted as a probabilistic status quo that is (pure) Nash equilibrium of a probabilised game. The Cartesian product structure enables a static approach. CP and BR approach is dynamic because a CP or BR equilibrium can be interpreted as a limit set of states that are tied together by explicit forces. It may be interesting to mix features from both approaches, and to present for instance a discrete and static notion of equilibrium. Actually, such an approach was already adopted in~\cite{SLR06}, whose purpose was to provide sequential tree games with a notion of discrete non deterministic equilibrium. This approach assumed partially ordered payoffs, and simple "backward induction" guarantees existence of non deterministic subgame perfect equilibrium. This result is superseded by~\cite{SLR2007-18} which adopts a completely different approach, but the discrete non determinism spirit can be further exploited.

\subsection{Contribution}

This paper introduces the concept of abstract strategic games, which corresponds to traditional strategic games where real-valued payoff functions have been replaced with abstract objects called outcomes. In addition, the usual total order over the reals has been replaced with binary relations, one per agent, that account for agent's preferences over the outcomes. Abstract strategic games thus generalise strategic games like abstract sequential tree games generalise sequential tree games. A notion of Nash equilibrium is defined, but not all abstract strategic games have a Nash equilibrium since traditional strategic games already lack this property.

Like Nash did for traditional strategic games, an attempt is made to introduce probabilities into these new games. However, it is mostly a failure because there does not seem to exist any extension of a poset to its barycentres that is relevant to the purpose. So, instead of saying that "an agent chooses a given strategy with some probability", this paper proposes to say that "the agent may choose the strategy", without further specification.

The discrete non determinism proposed above is implemented in the notion of non deterministic best response (ndbr ) multi strategic game. As hinted by the terminology, the best response approach is preferred over the convertibility preference approach for this specific purpose. (Note that discrete non determinism for abstract strategic games can be implemented in a formalism that is more specific and simpler than ndbr multi strategic games, but this general formalism will serve further purposes.) This paper defines the notion of ndbr equilibrium in these games, and a pre-fixed point result helps prove a sufficient condition for every ndbr multi strategic game to have an ndbr equilibrium. An embedding of abstract strategic games into ndbr multi strategic games provides abstract strategic games with a notion of non deterministic (nd ) equilibrium that generalises the notion of Nash equilibrium. Since every abstract strategic game has an nd equilibrium (under some condition), the discrete non deterministic approach succeeds where the probabilistic approach fails, \textit{i.e.} is irrelevant. This new approach lies between Nash's approach, which is continuous and static, and the abstract approaches of CP and BR games, which are discrete and dynamic. Indeed, this notion of nd equilibrium is discrete and static. It is deemed static because it makes use of the Cartesian product structure, which allows interpreting an equilibrium as a "static state of the game".

This paper also defines the notion of multi strategic game that is very similar to the notion of ndbr multi strategic game, while slightly less abstract. multi strategic games are actually a generalisation of both abstract strategic games and sequential graph games. Informally, they are games where a strategic game takes place at each node of a graph. (A different approach to "games network" can be found in~\cite{Manceny06}) They can thus model within a single game both sequential and simultaneous decision-making mechanisms. An embedding of multi strategic games into ndbr multi strategic games provides multi strategic games with a notion of non deterministic (nd ) equilibrium. In addition, a numerical example shows that the constructive proof of nd equilibrium existence can serve as a recommmendation to agents on how to play, while the notion of Nash equilibrium, as its stands, cannot lead to any kind of recommendation.

\subsection{Contents}

Section~\ref{sect:asg} defines abstract strategic games and their abstract Nash equilibria. Section~\ref{sect:cdnd} considers probabilities to relax the definition of Nash equilibrium in abstract strategic games, and concludes that discrete non determinism is required. Section~\ref{sect:fpfp} proves a pre-fixed point result. Section~\ref{sect:ndbrmsg} introduces the non deterministic best response multi strategic games and their non deterministic best response equilibria. Then it gives a sufficient condition for these games to have such an equilibrium. Section~\ref{sect:dsceq-asg} embeds abstract strategic games into ndbr multi strategic games, and thus provides a notion of (existing) non deterministic equilibrium for abstract strategic games. It also gives a few examples. Section~\ref{sect:dsceq-msg} defines multi strategic games and embeds them into ndbr multi strategic games, and thus provides a notion of non deterministic equilibrium for multi strategic games.

\section{Abstract Strategic Games}\label{sect:asg}

This section defines abstract strategic games and their abstract Nash equilibria. Two embeddings show that abstract strategic games can be seen as either CP games or BR games when preferences are acyclic.

Informally, abstract strategic games are traditional strategic games where real-valued payoff functions have been replaced with abstract objects named outcomes. In addition for each agent, one binary relation over outcomes accounts for the preference of the agent for some outcomes over some others.

\begin{defn}[Abstract strategic games] Abstract strategic games are 4-tuples $\quadruple{\Agent}
    { S}{P}{(\aPrefFct{}{\agent})_{\agent\in\Agent}}$ where:
\begin{itemize}
\item $\Agent$ is a non-empty set of agents,
\item $ S=\bigotimes_{\agent\in\Agent} S_{\agent}$
is the Cartesian product of non-empty sets of individual strategies, 
\item $P\,:\, S\to Oc$ is a function mapping strategy profiles to outcomes.
\item  $\aPrefFct{}{\agent}$ is a binary relation over outcomes, and $\aPref{}{\agent}{oc}{oc'}$ says that agent $a$ prefers $oc'$ over $oc$.
\end{itemize}
\end{defn}

The example below shows a traditional strategic game on the left-hand side, and an abstract strategic game on the right-hand side.

\[\begin{array}{c@{\hspace{2cm}}c}
         \begin{array}{c|c@{\quad}c@{\;\vline\;}c@{\quad}c|}
          \multicolumn{1}{c}{}&
	  \multicolumn{2}{c}{h_1}&
	  \multicolumn{2}{c}{h_2}\\
	  \cline{2-5}
 	  v_1 & 0 & 2 & 1 & 1\\
	  \cline{2-5}
	  v_2 & 0 & 1 & 2 & 0\\
	  \cline{2-5}
	 \end{array}
&
         \begin{array}{c|c@{\;\vline\;}c|}
          \multicolumn{1}{c}{}&
	  \multicolumn{1}{c}{h_1}&
	  \multicolumn{1}{c}{h_2}\\
	  \cline{2-3}
 	  v_1 & oc_1 & oc_2\\
	  \cline{2-3}
	  v_2 & oc_3 & oc_4\\
	  \cline{2-3}
	 \end{array}
\end{array}\]
\medskip

For a given game, each agent can compare strategy profiles by comparing their outcomes using the function $P$.

\begin{nota}
Let $\quadruple{\Agent}{ S}{P}{(\aPrefFct{}{\agent})_{\agent\in\Agent}}$ be an abstract strategic game and let $s$ and $s'$ be in $ S$. 
\[\aPref{}{a}{\simS}{\simS'}\quad\eqdef\quad\aPref{}{a}{P(\simS)}{P(\simS')}\]
\end{nota}

Happiness of an agent is defined below, in a convertibility preference style.

\begin{defn}[Agent happiness] Let $\quadruple{\Agent}
    { S}{P}{(\aPrefFct{}{\agent})_{\agent\in\Agent}}$ 
\[Happy(a,\simS)\quad\eqdef\quad\forall\simS'\in S,\,\neg(\simS'_{- a}=\simS_{- a}\,\wedge\,\aPref{}{a}{\simS}{\simS'})\]
\end{defn}

As usual, Nash equilibrium means happiness for all agents.

\begin{defn}[Nash Equilibrium\label{defn:spne}] 
\[\EqFct{}{g}{s}\quad\eqdef\quad\forall a\in\Agent,\,Happy(a,\simS)\]
\end{defn}

There exists a natural embedding of abstract strategic games into CP games, as described below.

\begin{lem}\label{lem:asg-cpg} Let $g=\quadruple{\Agent}{ S}{P}{(\aPrefFct{}{\agent})_{\agent\in\Agent}}$ be a strategic game. 
\[\aConvEq{}{\agent}{ s}{ s'}\quad\eqdef\quad s_{-\agent}= s'_{-\agent},\]
Then $g'=\quadruple{\Agent}{ S}{(\aConvEqFct{}{\agent})_{\agent\in\Agent}}{(\aPrefFct{}{\agent})_{\agent\in\Agent}}$ is a CP game and the embedding preserves and reflects Nash equilibria.
\[\EqFct{}{g}{s}\quad\Leftrightarrow\quad\EqFct{}{g'}{s}\]
\end{lem}

When agents' preferences are acyclic, there exists also a natural embedding of abstract strategic games into BR games, as described below.

\begin{lem}\label{lem:asg-brg} Let $g=\quadruple{\Agent}{ S}{P}{(\aPrefFct{}{\agent})_{\agent\in\Agent}}$ be a strategic game. Assume that the $\aPrefFct{}{\agent}$ are acyclic.
\[BR_a(s)\quad\eqdef\quad \{s_{-a}\}\times\{s'\in S_a\,\mid\,\forall s''\in S_a,\,\neg(\aPref{}{a}{s_{-a};s'}{s_{-a};s''})\}\]
Then $g'=\triple{\Agent}{ S}{(BR_a)_{\agent\in\Agent}}$ is a BR game and the embedding preserves and reflects Nash equilibria.
\[\EqFct{}{g}{s}\quad\Leftrightarrow\quad\EqFct{}{g'}{s}\]
\end{lem}

\section{From Continuous to Discrete non determinism}\label{sect:cdnd}

This section tries to apply Nash's probabilistic compromise to an instance of abstract strategic games. Facing a half-failure, it notices that continuous non determinism, \textit{i.e.} probabilities, carry "too much" information. Indeed, only a notion of discrete non deterministic strategies is needed to characterise the probabilistic Nash equilibria of the example. These non deterministic strategies are defined as non-empty subsets of strategies. 

Consider the following abstract strategic game involving agents $v$ and $h$. The game has no (pure) Nash equilibrium.

\[\begin{array}{c@{\hspace{2cm}}c}
\begin{array}{c@{\hspace{.5cm}\aPrefFct{}{V}\hspace{.5cm}}c@{\hspace{2cm}}c@{\hspace{.5cm}\aPrefFct{}{H}\hspace{.5cm}}c}
 oc_1 & oc_3 & oc_3 & oc_4 \\
 oc_4 & oc_2 & oc_2 & oc_1
\end{array}

&
\begin{array}{c|c@{\;\vline\;}c|}
          \multicolumn{1}{c}{}&
	  \multicolumn{1}{c}{h_1}&
	  \multicolumn{1}{c}{h_2}\\
	  \cline{2-3}
 	  v_1 & oc_1 & oc_2\\
	  \cline{2-3}
	  v_2 & oc_3 & oc_4\\
	  \cline{2-3}
\end{array}
\end{array}\]

Mixed strategies for abstract strategic games are defined the same way they are defined for traditional strategic games, \textit{i.e.} through probability distributions. Mixed strategy profiles are of the following form, where $\alpha$ and $\beta$ are probabilities that are chosen by agent $v$ and agent $h$ respectively.

\[\alpha\beta(v_1,h_1)+\alpha(1-\beta)(v_1,h_2)+(1-\alpha)\beta(v_2,h_1)+(1-\alpha)(1-\beta)(v_2,h_2)\]

In the traditional setting, this yields expected payoff functions. Since payoffs and probabilities are both real numbers, the expected payoffs are also real numbers. It is therefore natural to compare them by using the usual total order over the reals, \textit{i.e.} the same order that is used when comparing payoffs of pure strategy profiles. In the abstract setting though, mixing the strategies induces a new type of object, say "expected outcomes". These expected outcomes are objects of the following form, where $\alpha$ and $\beta$ are probabilities that are chosen by agent $v$ and agent $h$ respectively.

\[\alpha\beta oc_1+\alpha(1-\beta)oc_2+(1-\alpha)\beta oc_3+(1-\alpha)(1-\beta)oc_4\]

These new objects are not outcomes a priori, so there is no obvious way to compare them a priori. The most natural way may be the following one. When both $\alpha$ and $\beta$ are either $0$ or $1$, the expected outcome looks like an outcome. For instance, if $\alpha$ and $\beta$ equal $1$, the expected outcome is as follows.

\[1.oc_1+0.oc_2+0.oc_3+0.oc_4\]

Along the two preference relations that are defined in the strategic game above, it is possible to define preferences among these four specific expected outcomes. For instance, $\aPref{}{v}{oc_1}{oc_3}$ yields the following. 

\[1.oc_1+0.oc_2+0.oc_3+0.oc_4\quad\aPrefFct{}{v}\quad0.oc_1+0.oc_2+1.oc_3+0.oc_4\]

When either $\alpha$ or $\beta$ is either $0$ or $1$, the new preference relations can be extended naturally, \textit{i.e.} consistently with the original preferences. Informally, if agent $v$ prefers $oc_3$ over $oc_1$ then, by extension, he will prefer mixed outcomes giving more weight to $oc_3$ than to $oc_1$. For instance, if $\alpha'<\alpha$ then the preference of agent $v$ is extended as follows.

\[\alpha oc_1+0.oc_2+(1-\alpha)oc_3+0.oc_4\quad\aPrefFct{}{v}\quad\alpha' oc_1+0.oc_2+(1-\alpha')oc_3+0.oc_4\]

However, extending non-trivially the preferences to all expected outcomes would require an artificial choice. Indeed, consider the following two expected outcomes obtained by $\alpha=\frac{1}{2}$ and $\beta=\frac{1}{2}$ for the first one, and by $\alpha=\frac{1}{2}$ and $\beta=\frac{1}{3}$ for the second one.

\[\frac{1}{4}oc_1+\frac{1}{4}oc_2\quad+\quad\frac{1}{4}oc_3+\frac{1}{4}oc_4\]
\[\frac{1}{6}oc_1+\frac{1}{3}oc_2\quad+\quad\frac{1}{6}oc_3+\frac{1}{3}oc_4\]

In the strategic game above, these expected outcomes correspond to mixed strategy profiles that agent $h$ can convert to each other. On the one hand, the weights attributed to $oc_1$ and $oc_2$ are better in the first expected outcome, according to $h$. On the other hand, the weights attributed to $oc_3$ and $oc_4$ are better in the second expected outcome, according to $h$. These two arguments sound "contradictory". Moreover, there is nothing in the original preference relation that suggests to give priority to one argument over the other. So it is reasonable to say that $h$ prefers neither of these expected outcomes. Thus are defined the extensions of the preference relations. This completes the definition of the (probabilistic) abstract strategic game derived from the finite abstract strategic game example above. In such a setting, the probabilistic Nash equilibria are the mixed strategy profiles $\alpha\beta(v_1,h_1)+\alpha(1-\beta)(v_1,h2)+(1-\alpha)\beta(v_2,h_1)+(1-\alpha)(1-\beta)(v_2,h_2)$, where $0<\alpha<1$ and $0<\beta<1$. Therefore almost all mixed strategy profiles are probabilistic Nash equilibria, which does not seem not be a desirable property.

In the probabilistic setting, a strategy of an agent is said to be "used" if the agent gives a non-zero probability to this strategy. With this terminology, the above remark can be rephrased as follows. In the above strategic game with its above probabilistic extension, a mixed strategy profile is a probabilistic Nash equilibrium \textit{iff} both agents use both their strategies. This suggests that, in abstract strategic games, the actual values of the probabilities are irrelevant to the Nash equilibrium predicate. Only their being zero or not is relevant. This motivates the definition of discrete non deterministic strategies that only says which strategies are used. Note that an agent must use at least one strategy, like in the probabilistic (and the pure) setting. For each agent, a discrete non deterministic strategy can therefore be seen as a non-empty subset of the set of its strategies.

\section{A Simple Pre-Fixed Point Result}\label{sect:fpfp}

This section proves a pre-fixed-point result, \textit{i.e.} the existence of a $y$ such that $y\preceq F(y)$ for all $F$ and $\preceq$ that comply with given constraints.

Meet semi-lattices are defined below like in the literature. They are posets that guarantee existence of greatest lower bound of any two elements. The terminology of "meet" seems to come from the set-theoretic intersection.

\begin{defn}[Meet semi-lattice]
A meet semi-lattice is a partially ordered set $(S,\preceq)$, \textit{i.e.} the binary relation $\preceq$ is reflexive and transitive, such that any sets with two elements has a greatest lower bound.
\end{defn}

Defined as a specific type of posets, meet semi-lattices have algebraic properties. (Actually, meet semi-lattices are sometimes defined as algebraic structures from where an ordering is derived.)

\begin{defn}
In a partially ordered set, a greatest lower bound of two elements is unique, which induces the binary operator "greatest lower bound". This operator is commutative and associative, which enables the definition of the greatest lower bound of any non-empty finite subset of $S$. Let us call $inf$ this greatest lower bound function.
\end{defn}

Given a function from a meet semi-lattice to itself, a meeting point is an element of the lattice such that every decreasing sequence that starts with the meeting point is not too much "scattered" by the function. This is accurately described below.

\begin{defn}
Let $(S,\preceq)$ be a meet semi-lattice with least element $m$, and let $inf$ be the infimum function corresponding to $\preceq$. Let $F$ be a function from $S$ to itself and let $x$ be in $S$. Assume that for all $x_1\dots x_n$ such that $m\neq x_1\preceq\dots\preceq x_n\preceq x$, we have $inf(F(x_1),\dots,F(x_n),x)\neq m$. If $x\neq m$ then $x$ is said to be a $F$-meeting point, and one writes $M_F(x)$.
\end{defn}

The $F$-meeting point predicate is preserved by the greatest lower bound operator used with the image of the point by $F$, as stated below.

\begin{lem}\label{lem:mp-pres}
Let $(S,\preceq)$ be a meet semi-lattice with least element $m$; let $inf$ be the infimum function corresponding to $\preceq$; and let $F$ be a function from $S$ to itself. The following formula holds.
\[M_F(x)\,\Rightarrow\, M_F\circ inf(x,F(x))\]
\end{lem}

\begin{proof}
Assume $M_F(x)$, so $x$ and $m$ are different. By reflexivity $x\preceq x$, so $inf(x,F(x))\neq m$ by definition of meeting point. Assume $m\neq x_1\preceq\dots\preceq x_n\preceq inf(x,F(x))$, so $m\neq x_1\preceq\dots\preceq x_n\preceq x\preceq x$ since $inf(x,F(x))\preceq x$ by definition of $inf$. So $inf(F(x_1),\dots,F(x_n),F(x),x)\neq m$ since $x$ is a $F$-meeting point. Therefore $inf(F(x_1),\dots,F(x_n),inf(F(x),x))\neq m$, by associativity of the greatest lower bound operator underlying the infimum function $inf$. So $inf(x,F(x))$ is a $F$-meeting point.
\end{proof}

The $F$-meeting point predicate preservation can be combined with the assumption that there exists no infinite strictly decreasing sequence. In this case, iteration of lemma~\ref{lem:mp-pres} yields a non-trivial pre fixed point of $F$.

\begin{lem}\label{lem:pre-fx-pt}
Let $(S,\preceq)$ be a meet semi-lattice with least element $m$, and assume that $\preceq$ is well-founded. Let $F$ be a function from $S$ to itself. If there exists a $F$-meeting point, then there exists a $F$ pre fixed point different from $m$, \textit{i.e.} there exists $y\neq m$ such that $y\preceq F(y)$. 
\end{lem}

\begin{proof}
Assume $M_F(x_0)$. An infinite sequence of elements of $S$ is built as follows. It starts with $x_0$, and it is gradually defined by induction. Assume $x_n,\dots,x_0$ such that $M_F(x_n)$, and $x_{k+1}=inf(x_k,F(x_k))$ for all $0\leq k<n$. Let $x_{n+1}=inf(x_n,F(x_n))$. By Lemma~\ref{lem:mp-pres}, $M_F(x_{n+1})$. By well-foundness assumption, there exists $n$ such that $x_{n+1}=x_n$, which means that $x_n=inf(x_n,F(x_n))$, so $x_n\preceq F(x_n)$. Moreover, $M_F(x_n)$ by construction of the sequence, so $x_n\neq m$. 
\end{proof}

\section[ndbr multi Strategic Games]{Non deterministic best Response multi Strategic Games}\label{sect:ndbrmsg}

Using the concept of discrete non deterministic strategy, this section defines non deterministic best response multi strategic games and their non deterministic Nash equilibria. This section also proves a sufficient condition for non deterministic Nash equilibrium existence in every non deterministic best response sequential graph game. These results will be used to guarantee existence of non deterministic equilibrium for abstract strategic games and sequential graph game.

Informally, non deterministic best response multi strategic games involve agents who play on several (abstractions of) strategic games at the same time. Agents' strategies are non deterministic, \textit{i.e.} on each game each agent has to choose one or more (pure) strategies among his available strategies. When all the opponents of an agent have chosen their non deterministic strategies on all games, a function tells agent $a$ what his non deterministic best responses are. These games are formally defined below.

\begin{defn}[Non deterministic best response multi strategic games\label{defn:brsgg}]  $\quad\\$ An ndbr multi strategic game is pair $\langle (S^i_a)^{i\in I}_{a\in\Agent},(BR_a)_{a\in\Agent}\rangle$ complying with the following.
\begin{itemize}
\item $I$ is a non-empty set of indices and $\Agent$ is a non-empty set of agents.
\item For all $a$ in $\Agent$, $BR_a$ is a function from $\Sigma_{-a}$ to $\Sigma_a$,\\ 
where $\Sigma=\bigotimes_{ a\in\Agent}\Sigma_a$ and $\Sigma_a=\bigotimes_{i\in I}\powerset{S^i_a}-\{\emptyset\}$ and $\Sigma_{-a}=\bigotimes_{ a'\in\Agent-\{a\}}\Sigma_{a'}$.
\end{itemize}
Elements of $\Sigma_a$ are called nd strategies for $a$, and elements of $\Sigma$ are called nd strategy profiles.
\end{defn}

Informally, an agent is happy with an nd strategy profile if its own nd strategy is included in its best responses against other agents' nd strategies. Agents' happiness is formally defined as follows. 

\begin{defn}[Happiness]
Let $g=\langle (S^i_a)^{i\in I}_{a\in\Agent},(BR_a)_{a\in\Agent}\rangle$ be an ndbr multi strategic game, and let $\sigma$ be in $\Sigma$.
\[Happy(\sigma,a)\quad\eqdef\quad\sigma_a\subseteq BR_a(\sigma_{-a})\]
\end{defn}

As usual, (non deterministic) Nash equilibrium amounts to happiness for all agents.

\begin{defn}[Non deterministic Nash equilibrium]
Let $g$ be an ndbr multi strategic game, $g=\langle (S^i_a)^{i\in I}_{a\in\Agent},(BR_a)_{a\in\Agent}\rangle$.
\[Eq_g(\sigma)\quad\eqdef\quad\forall a\in\Agent,\,Happy(\sigma,a)\]
\end{defn}

The individual best response functions can be combined into a collective best response function from the non deterministic profiles into themselves.

\begin{defn}[Combined best response]
Given an ndbr multi strategic game with $(BR_a)_{a\in\Agent}$ a family of agent best responses. The combined best response is a function from $\Sigma$ to itself defined as follows.
\[BR(\sigma)\,\eqdef\,\bigotimes_{a\in\Agent}BR_a(\sigma_{-a})\]
\end{defn}

Through the combined best response function, the non deterministic Nash equilibria are characterised below as nd profiles included in their images by the combined best response.

\begin{lem}
An ndbr equilibrium for $g=\langle (S^i_a)^{i\in I}_{a\in\Agent},(BR_a)_{a\in\Agent}\rangle$ is characterised as follows.
\[Eq_g(\sigma)\quad\Leftrightarrow\quad\sigma\subseteq BR(\sigma)\]
\end{lem}

Like in BR games, it is easy to define agents' strict happiness.

\begin{defn}[Strict happiness]
Let $g=\langle (S^i_a)^{i\in I}_{a\in\Agent},(BR_a)_{a\in\Agent}\rangle$ be an ndbr multi strategic game, and let $\sigma$ be in $\Sigma$.
\[Happy^+(\sigma,a)\quad\eqdef\quad\sigma_a=BR_a(\sigma_{-a})\]
\end{defn}

Then, (non deterministic) strict Nash equilibrium is defined as strict happiness for all agents.

\begin{defn}[Non deterministic strict Nash equilibrium]
Let $g$ be an ndbr multi strategic game, $g=\langle (S^i_a)^{i\in I}_{a\in\Agent},(BR_a)_{a\in\Agent}\rangle$, and let $\sigma$ be in $\Sigma$.
\[Eq^+_g(\sigma)\quad\eqdef\quad\forall a\in\Agent,\,Happy^+(\sigma,a)\]
\end{defn}

The following embedding of non deterministic best response multi strategic games into BR games preserves an reflects equilibria.

\begin{lem}
Let $g=\langle (S^i_a)^{i\in I}_{a\in\Agent},(BR_a)_{a\in\Agent}\rangle$ be an ndbr multi strategic game. Define $BR'_a$ with a Cartesian product below, for $\sigma\in\Sigma$.
\[BR'_a(\sigma)\quad\eqdef\quad\{\sigma_{-a}\}\times\bigotimes_{i\in I}\powerset{BR^i_a(\sigma_{-a})}-\{\emptyset\}\]
Where $BR^i_a(\sigma_{-a})$ is the $i$-projection of $BR_a(\sigma_{-a})$. Then the object $g'$ defined by $g'=\triple{\Agent}{\Sigma}{(BR'_a)_{a\in\Agent}}$ is a BR game and Nash equilibria correspond as follows.
\[Eq_{g'}(\sigma)\quad\Leftrightarrow\quad Eq_g(\sigma)\]
\end{lem}

\begin{proof}
Let $\sigma$ be in $\Sigma=\bigotimes_{ a\in\Agent}\bigotimes_{ i\in I}(\powerset{ S^i_{a}}-\{\emptyset\})$. The following chain of equivalences proves the claim. $Eq_{g'}(\sigma)\,\Leftrightarrow\,\forall a,\sigma\in BR'_a(\sigma)\,\Leftrightarrow\,\forall a,\sigma\in\{\sigma_{-a}\}\times\bigotimes_{i\in I}\powerset{BR^i_a(\sigma_{-a})}-\{\emptyset\}\,\Leftrightarrow\,\forall a,\sigma_a\in\bigotimes_{i\in I}\powerset{BR^i_a(\sigma_{-a})}-\{\emptyset\}\,\Leftrightarrow\,\forall a,\sigma_a\subseteq BR_a(\sigma_{-a})\,\Leftrightarrow\, Eq_{g}(\sigma)$
\end{proof}

The remainder of the section invokes the fixed-point results of section~\ref{sect:fpfp}, but prior to that, a meet lattice needs to be identified. 

\begin{lem}
Given an ndbr multi strategic game $\langle (S^i_a)^{i\in I}_{a\in\Agent},(BR_a)_{a\in\Agent}\rangle$, the poset $(\Sigma\cup\{\emptyset\},\subseteq)$ is a meet semi-lattice with least element $\emptyset$.
\end{lem}

A first equilibrium existence result is given below.

\begin{lem}\label{lem:s-mpt-ndne}
Given an ndbr multi strategic game $\langle (S^i_a)^{i\in I}_{a\in\Agent},(BR_a)_{a\in\Agent}\rangle$, if there exists a $BR$-meeting point, then there exists a non deterministic Nash equilibrium.
\end{lem}

\begin{proof}
By lemma~\ref{lem:pre-fx-pt}, if there exists a $BR$-meeting point, then there exists a non-empty pre fixed point $\sigma$ for $BR$, \textit{i.e.} there exists $\sigma\neq\emptyset$ such that $\sigma\subseteq BR(\sigma)$. It is an ndbr equilibrium by definition. 
\end{proof}

The main equilibrium existence result is stated below.

\begin{lem}\label{lem:ndne}
Consider an ndbr multi strategic game $\langle (S^i_a)^{i\in I}_{a\in\Agent},(BR_a)_{a\in\Agent}\rangle$. Let $\sigma$ be in $\Sigma$. Assume that for all agents $a$, for all $\gamma_1\dots\gamma_n$ in $\Sigma_{-a}$, if $\gamma_n\subseteq\dots\subseteq\gamma_1\subseteq\sigma_{-a}$ then $\cap_{1\leq k\leq n}BR_a(\gamma_k)\cap\sigma_a\neq\emptyset$. In this case, the game has an ndbr equilibrium.
\end{lem}

\begin{proof}
By lemma~\ref{lem:s-mpt-ndne}, it suffices to show that there exists a $BR$-meeting point. Let us prove that $\sigma$ is such a meeting point. First of all, $\sigma$ is non-empty since it belongs to $\Sigma$. Second, assume $\sigma^1\dots\sigma^n$ in $\Sigma$ such that $\sigma^1\subseteq\dots\subseteq\sigma^n\subseteq\sigma$. So for all agents $a$, $\sigma^1_{-a}\subseteq\dots\subseteq\sigma^n_{-a}\subseteq\sigma_{-a}$. By assumption, $\cap_{1\leq k\leq n}BR_a(\sigma^k_{-a})\cap\sigma_a\neq\emptyset$, which amounts to $(\cap_{1\leq k\leq n}BR(\sigma^k)\cap\sigma)_a\neq\emptyset$. Since this holds for all $a$, we have $\cap_{1\leq k\leq n}BR(\sigma^k)\cap\sigma\neq\emptyset$. Therefore $\sigma$ is a $BR$-meeting point.
\end{proof}

Equilibria are preserved when "increasing" the best response functions, as stated below.

\begin{lem}\label{lem:inc-pres-eq}
Let $g=\langle (S^i_a)^{i\in I}_{a\in\Agent},(BR_a)_{a\in\Agent}\rangle$ and $g'=\langle (S^i_a)^{i\in I}_{a\in\Agent},(BR'_a)_{a\in\Agent}\rangle$ be two ndbr multi strategic games such that for all $a$ in $\Agent$, for all $\gamma$ in $\Sigma_{-a}$, $BR_a(\gamma)\subseteq BR'_a(\gamma)$. In this case, the following implication holds.
\[\EqFct{}{g}{\sigma}\quad\Rightarrow\quad\EqFct{}{g'}{\sigma}\]
\end{lem}

\begin{proof}
Since for all $a$ in $\Agent$, for all $\gamma$ in $\Sigma_{-a}$, $BR_a(\gamma)\subseteq BR'_a(\gamma)$, it follows that for all $\sigma$ in $\Sigma$, $BR(\sigma)\subseteq BR'(\sigma)$. So, $\sigma\subseteq BR(\sigma)$ implies $\sigma\subseteq BR'(\sigma)$.
\end{proof}

\section[Equilibrium for Abstract Strategic Games]{Discrete and Static Compromising Equilibrium for Abstract Strategic Games}\label{sect:dsceq-asg}

Subsection~\ref{subsect:nd eq-ex} embeds abstract strategic games into ndbr multi strategic games, and thus provides a notion of non deterministic Nash equilibrium for abstract strategic games. Finally it proves equilibrium existence for all abstract strategic games. Subsection~\ref{subsect:ex} gives an example of such non deterministic Nash equilibrium on a strategic game with real-valued payoff functions. The building of the equilibrium is described step by step. Subsection~\ref{subsect:comp} suggests by a numerical example that the constructive proof of existence of such an equilibrium yields a notion of recommendation that is better on average than playing randomly.

\subsection{non deterministic Equilibrium Existence}\label{subsect:nd eq-ex}

The following definition extends orders over functions' codomains to orders over (restrictions of) functions. 

\begin{defn}
Let $f$ and $f'$ be functions of type $A\to B$, and let $A'$ be a subset of $A$. Let $\prec$ be an irreflexive binary relation over $B$, and let $\preceq$ be its reflexive closure. 
\begin{eqnarray*}
f\preceq^{A'} f'&\quad\eqdef\quad&\forall x\in A',\,f(x)\preceq f'(x)\\
f\prec^{A'} f'&\quad\eqdef\quad& f\preceq^{A'} f'\quad\wedge\quad\exists x\in A',\,f(x)\prec f'(x)
\end{eqnarray*}
One simply writes $f\prec f'$ instead of $f\prec^A f'$.
\end{defn}

For example if the two functions are represented by vectors of naturals with the usual total order, then $(1,1)<(1,2)$, $(0,2)<(1,2)$ and $(0,1)<(1,2)$. 

The extension above preserves strict partial orders, as stated below.

\begin{lem}
If $\prec$ is a strict partial order over $B$, then the derived $\prec^{A}$ over functions of type $A\to B$ is also a strict partial order.
\end{lem}

Given finitely many functions of the same type, given an order on the codomain, given finitely many subsets of the domain that are totally ordered by inclusion, one of these functions is maximal for the extension order with respect to each of the subsets. This is proved by the following.

\begin{lem}\label{lem:mutli-max}
Let $E$ be a finite set of functions of type $A\to B$. Let $\prec$ be an irreflexive and transitive binary relation over $B$, and let $\preceq$ be its reflexive closure. For any $\emptyset\neq A_0\subseteq\dots\subseteq A_n\subseteq A$ there exists an $f$ in $E$ that is maximal with respect to all the extended orders from $\prec^{A_0}$ to $\prec^{A_n}$.
\end{lem}

\begin{proof}
By induction on $n$. First case, $n=0$. There exists an $f$ in $E$ that is maximal, \textit{i.e.} have no successor, with respect to $\prec^{A_0}$ since $\prec^{A_0}$ is a partial order. Second case, assume that the claim holds for $n$ and let $\emptyset\neq A_0\subseteq\dots\subseteq A_{n+1}$ be subsets of $A$. Let $f$ in $E$ be maximal with respect to all the $\prec^{A_0}\dots\prec^{A_n}$. Let $f'$ be maximal with respect to $\prec^{A_{n+1}}$ among the functions in $E$ that coincide with $f$ on $A_{n}$. Since $\emptyset\neq A_0\subseteq\dots\subseteq A_{n}$, the function $f'$ is maximal in $E$ with respect to all the $\prec^{A_0}\dots\prec^{A_n}$. Let $f''$ in $E$ be such that $f'\preceq^{A_{n+1}}f''$. So $f'\preceq^{A_{n}}f''$ since $A_n\subseteq A_{n+1}$. Therefore $f'$ and $f''$ coincide on $A_n$ by maximality of $f'$ with respect to $\prec^{A_n}$. Since $f$ and $f'$ coincide on $A_n$, the functions $f$ and $f''$ also coincide on $A_n$. Therefore $\neg(f'\prec^{A_{n+1}}f'')$ by definition of $f'$, which shows that $f'$ is also maximal with respect to $\prec^{A_{n+1}}$.
\end{proof}

Below, an ndbr multi strategic game is built from an abstract strategic game. This ndbr multi strategic game always has an ndbr equilibrium. In the notation $\aPref{\gamma}{a}{s}{s'}$ below, the strategies $s$ and $s'$ are seen as functions from $ S_a$ to the outcomes.

\begin{lem}\label{lem:ndbr Nash-eq-ex}
Let $g=\quadruple{\Agent}{ S}{P}{(\aPrefFct{}{\agent})_{\agent\in\Agent}}$ be an abstract strategic game. Assume that the $\aPrefFct{}{\agent}$ are strict partial orders, \textit{i.e.} irreflexive and transitive. For each agent $a$ and each $\gamma$ in $\Sigma_{-a}$, the following defines a subset of $S_a$.
\begin{eqnarray*}
BR_a(\gamma)\quad\eqdef\quad\{s\in S_a\,\mid\, & \forall s'\in S_a,\,\neg(\aPref{S_{-a}}{a}{s}{s'}) & \quad\wedge\\
  & \forall s'\in S_a,\,\neg(\aPref{\gamma}{a}{s}{s'}) & \quad\wedge\\
  &\exists c\in\gamma,\forall s'\in S_a,\,\neg(\aPref{\{c\}}{a}{s}{s'}) & \quad\}
\end{eqnarray*}
The object $\langle ( S_a)_{a\in\Agent},(BR_a)_{a\in\Agent}\rangle$ is an ndbr multi strategic game, and it has an ndbr equilibrium. 
\end{lem}

\begin{proof}
Let $a$ be an agent. First note that, through the function $P$, every strategy $s$ in $S_a$ can be seen as a function from $S_{-a}$ to the outcomes $Oc$. Let $\gamma_1\subseteq\dots\subseteq\gamma_n$ be in $\Sigma_{-a}$ and let $c$ be in $\gamma_1$. According to lemma~\ref{lem:mutli-max}, there exists an $s$ in $S_a$ that is maximal with respect to $\aPrefFct{\{c\}}{a},\aPrefFct{\gamma_1}{a},\dots,\aPrefFct{\gamma_n}{a}, \aPrefFct{S_{-a}}{a}$. By definition of $BR_a$ (third conjunct), this strategy $s$ is in all the $BR_a(\gamma_1),\dots,BR_a(\gamma_n)$. First, this shows that $BR_a$ returns non-empty sets, so $\langle ( S_a)_{a\in\Agent},(BR_a)_{a\in\Agent}\rangle$ is indeed an ndbr multi strategic game. Second, this game has an ndbr equilibrium by lemma~\ref{lem:ndne}. 
\end{proof}

The ndbr equilibrium for the derived ndbr multi strategic game is called a non deterministic equilibrium for the original abstract strategic game. Other similar definitions are possible for non deterministic equilibrium. Especially definitions more generous than the one in lemma~\ref{lem:ndbr Nash-eq-ex}, which guarantee the existence of a non deterministic equilibrium, according to  lemma~\ref{lem:inc-pres-eq}. This yields the following lemma.

\begin{lem}\label{lem:multi ndbr eq}
Let $g=\quadruple{\Agent}{ S}{P}{(\aPrefFct{}{\agent})_{\agent\in\Agent}}$ be a strategic game. Assume that the $\aPrefFct{}{\agent}$ are strict partial orders, \textit{i.e.} irreflexive and transitive. For each agent $a$ and each $\gamma$ in $\Sigma_{-a}$, the following defines three subsets of $S_a$.
\begin{eqnarray*}
BR^1_a(\gamma)&\quad\eqdef\quad&\{s\in S_a\,\mid\, \forall s'\in S_a,\,\neg(\aPref{S_{-a}}{a}{s}{s'})\}\\
BR^2_a(\gamma)&\quad\eqdef\quad&\{s\in S_a\,\mid\,\forall s'\in S_a,\,\neg(\aPref{\gamma}{a}{s}{s'})\} \\
BR^3_a(\gamma)&\quad\eqdef\quad&\{s\in S_a\,\mid\,\exists c\in\gamma,\forall s'\in S_a,\,\neg(\aPref{\{c\}}{a}{s}{s'})\}\\
BR^4_a(\gamma)&\quad\eqdef\quad&\{s\in S_a\,\mid\,\forall s'\in S_a,\,\exists c\in\gamma,\, \neg(\aPref{\{c\}}{a}{s}{s'})\}
\end{eqnarray*}
The object $\triple{\Agent}{ S}{(BR^i_a)_{a\in\Agent}}$, for $i$ between $1$ and $4$, is an ndbr multi strategic game, and it has an ndbr equilibrium. 
\end{lem}

The first three $BR^i_a$ above correspond tho the three conjuncts of the $BR_a$ of lemma~\ref{lem:ndbr Nash-eq-ex}, and $BR^4_a$ is even more generous than $BR^2_a$. Note that $BR^2$ and $BR^4$ somewhat relate to the notions of dominated strategy, studied in~\cite{Gale53} and~\cite{LR57}, and rationalizability, studied in~\cite{Bernheim84} and~\cite{Pearce84}. These notions are more recently discussed in~\cite{HK02}, for instance. (These notions
are also related to "backward induction", but this thesis does not further explore the matter.)

The rest of the subsection discusses a few properties of these equilibria.

\begin{defn}[Cartesian union]
Let $\bigotimes_{i\in\ I}A_i$ be a cartesian product. The Cartesian union is defined as follows within $\bigotimes_{i\in\ I}A_i$.
\[p_i(D)\quad\eqdef\quad\{x_i\,\mid\,x\in D\}\]
\[B\cup^\times C\quad\eqdef\quad\bigotimes_{i\in\ I}p_i(B)\cup p_i(C)\]
Where $p_i$ is the projection on $A_i$.
\end{defn}

The equilibria related to $BR^1$ define a simple structure, as stated below.

\begin{lem}
The ndbr equilibria related to $BR^1$ are the elements of $\Sigma$ that are included in $\bigotimes_{a\in\Agent}\{s\in S_a\,\mid\,  \forall s'\in S_a,\,\neg(\aPref{S_{-a}}{a}{s}{s'})\}$. 
\end{lem}

\begin{proof}
For all $\sigma$ in $\Sigma$, $BR^1(\sigma)=\bigotimes_{a\in\Agent}\{s\in S_a\,\mid\,  \forall s'\in S_a,\,\neg(\aPref{S_{-a}}{a}{s}{s'})\}$.
\end{proof}

The following lemma states that the ndbr equilibria related to $BR^3$ define a Cartesian union lattice.

\begin{lem}
Let $g=\quadruple{\Agent}{ S}{P}{(\aPrefFct{}{\agent})_{\agent\in\Agent}}$ be a strategic game. Assume that the $\aPrefFct{}{\agent}$ are strict partial orders, \textit{i.e.} irreflexive and transitive. If equilibrium is defined through $BR^3$ then the following holds.
\[Eq_g(\sigma)\quad\wedge\quad Eq_g(\sigma')\quad\Rightarrow\quad Eq_g(\sigma\cup^\times\sigma')\]
\end{lem}

\begin{proof}
It suffices to prove that $\sigma\cup^\times\sigma'\subseteq BR^3(\sigma\cup^\times\sigma')$. Let $s$ be in $\sigma\cup^\times\sigma'$. If $s$ is in $\sigma$ then it is also in $BR^3(\sigma)$ by equilibrium assumption. So for every agent $a$, there exists $c$ in $\sigma_{-a}$ (so $c$ is also in $\sigma_{-a}\cup^\times\sigma'_{-a}$) such that for all $s'$ in $S_a$, $neg(\aPref{\{c\}}{a}{s}{s'})$. Therefore $s$ is in $BR^3(\sigma\cup^\times\sigma')$. Same scenario if $s$ is in $\sigma'$. So $\sigma\cup^\times\sigma'\subseteq BR^3(\sigma\cup^\times\sigma')$.
\end{proof}

The following lemma states that the ndbr equilibria related to $BR^4$ define a Cartesian union lattice.

\begin{lem}
Let $g=\quadruple{\Agent}{ S}{P}{(\aPrefFct{}{\agent})_{\agent\in\Agent}}$ be a strategic game. Assume that the $\aPrefFct{}{\agent}$ are strict partial orders, \textit{i.e.} irreflexive and transitive. If equilibrium is defined through $BR^4$ then the following holds.
\[Eq_g(\sigma)\quad\wedge\quad Eq_g(\sigma')\quad\Rightarrow\quad Eq_g(\sigma\cup^\times\sigma')\]
\end{lem}

\begin{proof}
It suffices to prove that $\sigma\cup^\times\sigma'\subseteq BR^4(\sigma\cup^\times\sigma')$. Let $s$ be in $\sigma\cup^\times\sigma'$. If $s$ is in $\sigma$ then it is also in $BR^4(\sigma)$ by equilibrium assumption. So for all $s'$ in $S_a$, there exists $c$ in $\sigma_{-a}$ such that $\neg(\aPref{\{c\}}{a}{s}{s'})$, and each of these $c$ also belongs to $\sigma_{-a}\cup^\times\sigma'_{-a}$. Therefore $s$ is in $BR^4(\sigma\cup^\times\sigma')$. Same scenario if $s$ is in $\sigma'$. So $\sigma\cup^\times\sigma'\subseteq BR^4(\sigma\cup^\times\sigma')$.
\end{proof}

\subsection{Example}\label{subsect:ex}

The proof of existence of an ndbr equilibrium is constructive, so it provides for free an algorithm that computes such an equilibrium. The time complexity of the algorithm is polynomial with respect to the number of strategy profiles $|S|$ (when each agent has at least two available strategies). Indeed informally, each call to $BR$ dismisses at least one strategy of one agent, so it dismisses at least one profile. Therefore $BR$ is called at most $|S|$ times. Each use of $BR$ invokes all the $BR_a$, which needs (at most) to consider each agent strategy and decide whether or not this strategy is a best response. So there are at most $|S|$ such decisions. Such a decision requires at most $3\times|S|$ calls to a $\aPrefFct{}{a}$. Therefore the time complexity of finding an equilibrium is at most cubic in the number of profiles $|S|$. This is a very rough approximation whose only purpose is to show that complexity is polynomial.

The example below is a strategic game with natural-valued payoff functions. Preference between naturals invokes the usual total order over the naturals. Let us apply the equilibrium algorithm to the game.

\[\begin{array}{c|c@{\;}c@{\;\vline\;}c@{\;}c@{\;\vline\;}c@{\;}c@{\;\vline\;}c@{\;}c@{\;\vline\;}c@{\;}c|}
    \multicolumn{1}{c}{}&
    \multicolumn{2}{c}{h_1}&
    \multicolumn{2}{c}{h_2}&
    \multicolumn{2}{c}{h_3}&
    \multicolumn{2}{c}{h_4}&
    \multicolumn{2}{c}{h_5}\\
    \cline{2-11}
    v_1 & 0 & 0 & 3 & 2 & 2 & 2 & 2 & 1 & 3 & 0 \\
    \cline{2-11}
    v_2 & 3 & 3 & 0 & 2 & 0 & 0 & 2 & 1 & 3 & 2 \\
    \cline{2-11}
    v_3 & 2 & 2 & 1 & 0 & 0 & 1 & 3 & 1 & 0 & 2 \\
    \cline{2-11}
    v_4 & 1 & 0 & 1 & 2 & 1 & 2 & 1 & 2 & 1 & 0 \\
    \cline{2-11}
    v_5 & 2 & 0 & 1 & 0 & 0 & 0 & 0 & 1 & 0 & 0 \\
    \cline{2-11}
  \end{array}\]

Informally, $V$ may "play" either $v_1$ or $v_2$ or $v_3$ or $v_4$ or $v_5$. In that context, column $h_5$ is smaller than $h_1$ according to agent $H$, so $h_5$ is not a best response of agent $H$. In the same way row $v_5$ is smaller than row $v_3$. In addition, row $v_4$ is not a best response of agent $V$ because for each column $h_i$, row $v_4$ is smaller than some other row. Therefore, the combined best responses for the whole game are rows $v_1$ to $v_3$ and columns $h_1$ to $h_4$, as shown below.  

\[\begin{array}{c|c@{\;}c@{\;\vline\;}c@{\;}c@{\;\vline\;}c@{\;}c@{\;\vline\;}c@{\;}c|}
    \multicolumn{1}{c}{}&
    \multicolumn{2}{c}{h_1}&
    \multicolumn{2}{c}{h_2}&
    \multicolumn{2}{c}{h_3}&
    \multicolumn{2}{c}{h_4}\\
    \cline{2-9}
    v_1 & 0 & 0 & 3 & 2 & 2 & 2 & 2 & 1 \\
    \cline{2-9}
    v_2 & 3 & 3 & 0 & 2 & 0 & 0 & 2 & 1 \\
    \cline{2-9}
    v_3 & 2 & 2 & 1 & 0 & 0 & 1 & 3 & 1  \\
    \cline{2-9}
  \end{array}\]

Now, agent $V$ may play either $v_1$ or $v_2$ or $v_3$. Column $h_4$ is not a best response of agent $H$ because for each row $v_1$ to $v_3$, column $h_4$ is smaller than some other column. So the game "shrinks" again as show below.

\[\begin{array}{c|c@{\;}c@{\;\vline\;}c@{\;}c@{\;\vline\;}c@{\;}c|}
    \multicolumn{1}{c}{}&
    \multicolumn{2}{c}{h_1}&
    \multicolumn{2}{c}{h_2}&
    \multicolumn{2}{c}{h_3}\\
    \cline{2-7}
    v_1 & 0 & 0 & 3 & 2 & 2 & 2 \\
    \cline{2-7}
    v_2 & 3 & 3 & 0 & 2 & 0 & 0 \\
    \cline{2-7}
    v_3 & 2 & 2 & 1 & 0 & 0 & 1 \\
    \cline{2-7}
  \end{array}\]

In the same way, for columns $h_1$ to $h_3$, row $v_3$ is smaller than some other row.

\[\begin{array}{c|c@{\;}c@{\;\vline\;}c@{\;}c@{\;\vline\;}c@{\;}c|}
    \multicolumn{1}{c}{}&
    \multicolumn{2}{c}{h_1}&
    \multicolumn{2}{c}{h_2}&
    \multicolumn{2}{c}{h_3}\\
    \cline{2-7}
    v_1 & 0 & 0 & 3 & 2 & 2 & 2 \\
    \cline{2-7}
    v_2 & 3 & 3 & 0 & 2 & 0 & 0 \\
    \cline{2-7}
  \end{array}\]

Column $h_3$ is smaller than column $h_2$. This yield the following irreducible game.

\[\begin{array}{c|c@{\;}c@{\;\vline\;}c@{\;}c|}
    \multicolumn{1}{c}{}&
    \multicolumn{2}{c}{h_1}&
    \multicolumn{2}{c}{h_2}\\
    \cline{2-5}
    v_1 & 0 & 0 & 3 & 2 \\
    \cline{2-5}
    v_2 & 3 & 3 & 0 & 2 \\
    \cline{2-5}
  \end{array}\]

Therefore $\{v_1,v_2\}\times\{h_1,h_2\}$ is a non deterministic equilibrium for the original game.

\subsection{Comparisons}\label{subsect:comp}

Consider the following class of games $G$, where $*$ can take two values, namely $1$ and $-1$, and where the preferences are along the usual order $-1<1$.

\[\begin{array}{c|c@{\;}c@{\;\vline\;}c@{\;}c|}
    \multicolumn{1}{c}{}&
    \multicolumn{2}{c}{h_1}&
    \multicolumn{2}{c}{h_2}\\
    \cline{2-5}
    v_1 & * & * & * & * \\
    \cline{2-5}
    v_2 & * & * & * & * \\
    \cline{2-5}
  \end{array}\]

For each agent, the arithmetic mean of its payoff over the four payoff functions of all games in the class is $0$, by a "simple symmetry". However
for each agent, the arithmetic mean of its payoff over the payoff functions inside the ndbr equilibrium of all games in the class is $3/8$.

\begin{lem}
For a game $g$ in $G$, let $eq(g)$ be the ndbr equilibrium built by the proof of lemma~\ref{lem:ndbr Nash-eq-ex}.
\[\frac{1}{|G|}\times\sum_{g\in G}\frac{1}{|e(g)|}\times\sum_{s\in eq(g)}P(s,v)=\frac{3}{8}\]
\end{lem}

\begin{proof}
In the games of class $G$, considering only the payoffs of one agent yields matrices like below. The first row displays the matrices whose two rows are equivalent to agent $v$ (the agent choosing the row of the matrix). The second row displays the matrices whose two rows are not equivalent to agent $v$. The figure on the top of matrices represent how many actual matrices they represent up to rotation. 
 
\[\begin{array}{c@{\hspace{1cm}}c@{\hspace{1cm}}c@{\hspace{1cm}}c}
1&2&2&1\\

\begin{array}{|c@{\;\vline\;}c|}
    \cline{1-2}
    1 & 1 \\
    \cline{1-2}
    1 & 1 \\
    \cline{1-2}
  \end{array}
&
\begin{array}{|c@{\;\vline\;}c|}
    \cline{1-2}
    1 & -1 \\
    \cline{1-2}
    1 & -1 \\
    \cline{1-2}
  \end{array}
&
\begin{array}{|c@{\;\vline\;}c|}
    \cline{1-2}
    1 & -1 \\
    \cline{1-2}
    -1 & 1 \\
    \cline{1-2}
  \end{array}
&
\begin{array}{|c@{\;\vline\;}c|}
    \cline{1-2}
    -1 & -1 \\
    \cline{1-2}
    -1 & -1 \\
    \cline{1-2}
  \end{array}\\\\

4&2&4\\

\begin{array}{|c@{\;\vline\;}c|}
    \cline{1-2}
    1 & 1 \\
    \cline{1-2}
    1 & -1 \\
    \cline{1-2}
  \end{array}
&
\begin{array}{|c@{\;\vline\;}c|}
    \cline{1-2}
    1 & 1 \\
    \cline{1-2}
    -1 & -1 \\
    \cline{1-2}
  \end{array}
&
\begin{array}{|c@{\;\vline\;}c|}
    \cline{1-2}
    1 & -1 \\
    \cline{1-2}
    -1 & -1 \\
    \cline{1-2}
  \end{array}
\end{array}\]

The payoff that is induced by the first matrix is $1$: whatever may be the matrix of agent $h$, agent $v$ gets $1$. In the same way, the fourth matrix induces $-1$. Now assume that the payoff matrix of agent $v$ is the second one. When the payoff matrix of agent $h$ ranges over the matrices above, $eq(g)$ will involve sometimes only the left-hand side, sometimes only the right-hand side and sometimes (the same number of times by symmetry), and sometimes both. So on average, the third matrix yields payoff $0$. Same scenario for the third matrix.

The payoff that is induced by the sixth matrix is $1$. The fifth matrix induces payoff $1$ too, because an equilibrium involves only the first row whatever the matrix of agent $h$ is. As to the seventh matrix it induces payoff $0$ "by symmetry", because an equilibrium involves only the first row whatever the matrix of agent $h$ is. Therefore the arithmetic mean is $\frac{1+4+2-1}{16}=\frac{3}{8}$.
\end{proof}

The rest of the subsection establishes a connection between the non deterministic equilibria of abstract strategic games and the or-best response strict equilibria of abstract strategic games seen as BR games.

\begin{lem}
Let $g=\quadruple{\Agent}{ S}{P}{(\aPrefFct{}{\agent})_{\agent\in\Agent}}$ be an abstract strategic game. Assume that the $\aPrefFct{}{\agent}$ are strict partial orders, \textit{i.e.} irreflexive and transitive. For each agent $a$ and each $\gamma$ in $\Sigma_{-a}$, the following defines a subset of $S_a$.
\[BR^3_a(\gamma)\quad\eqdef\quad\{s\in S_a\,\mid\,\exists c\in\gamma,\forall s'\in S_a,\,\neg(\aPref{\{c\}}{a}{s}{s'})\}\]
The object $g^3=\triple{\Agent}{ S}{(BR^3_a)_{a\in\Agent}}$ is an ndbr multi strategic game.

For each agent $a$ and each $s$ in $ S$, the following defines a non-empty subset of $ S$, more specifically an element of $\Sigma$.
\[BR'_a(s)\quad\eqdef\quad \{s_{-a}\}\times\{s'\in S_a\,\mid\,\forall s''\in S_a,\,\neg(\aPref{\{s_{-a}\}}{a}{s'}{s''})\}\]
The object $g'=\triple{\Agent}{ S}{(BR'_a)_{a\in\Agent}}$ is a BR game. Then the following holds.
\[Eq^+_{g'}(C)\quad\Rightarrow\quad Eq_g(p(C))\]
Where $p(C)=\bigotimes_{a\in\Agent}p_a(C)$ is the smallest (for set inclusion) ndbr multi strategy profile including $C$.
\end{lem}

\begin{proof}
By lemma~\ref{lem:multi ndbr eq}, $g^3$ is an ndbr multi strategic game, and by lemma~\ref{lem:asg-brg}, $g'$ is a BR game. By assumption and by definition of or-best response strict equilibrium, $C=BR'^\cup(C)$. One must show that $p(C)\subseteq BR^3(p(C))$. Let $s$ be in $p(C)$. It suffices to prove that $s_a$ is in $BR_a^3(p_{-a}(C))$ for all $a$. Let $a$ be an agent. By definition of $p$, $s_a$ is in $p_a(C)$, and by definition of $p_a$ there exists $c$ in $ S_{-a}$ such that $c;s_a$ is in $C$. Such a $c$ is in $p_{-a}(C)$. Let $s_1\to_a s_2$ stand for $s_2\in BR'_a(s_1)$ and $\to$ stand for $\cup_{a\in\Agent}\to_a$. Case split on whether or not $c;s_a\to_a c;s_a$. If $c;s_a\to_a c;s_a$ then, by definition, $\neg\aPref{\{c\}}{a}{s_a}{s'})$ for all $s'$ in $S_a$, so $s_a$ is in $BR_a^3(p_{-a}(C))$. Now assume that $c;s_a\not\to_a c;s_a$. By definition, $BR'_a$ returns non-empty sets, so $c;s_a\not\to_a c;s'_a$ for some $s'_a\neq s_a$. Recall that $C$ is a $\to$ strongly connected component, so  $c;s'_a\to^+c;s_a$, where $\to^+$ is the transitive closure of $\to$. This decomposes in $c;s'_a\to^+c';s''_a\to_ac';s_a\to^+c;s_a$ for some $c'$ and $s''_a$. More specifically, $c';s''_a\to_ac';s_a$, with $c'$ in $p_{-a}(C)$. This implies that $s_a$ is in $BR_a^3(p_{-a}(C))$. 
\end{proof}

The ndbr equilibrium given by the implication $Eq^+_{g'}(C)\,\Rightarrow\, Eq_g(p(C))$ above is not always strict. More specifically, the following example shows that $Eq^+_{g'}(C)\,\wedge\,\neg Eq^+_g(p(C))$ for some or-best response equilibrium $C$. The example involves three agents each of whose converts along one dimension of the cube. Outcomes are payoff functions. The three nodes linked by doublelines constitute an or-best response strict equilibrium. The projection of the equilibrium is made of the four upper nodes, and the combined best response to this is the whole cube due to the two right-hand nodes in the back.

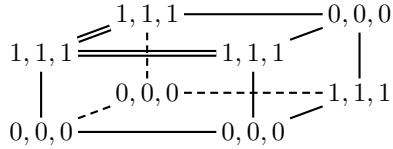
\[\psmatrix
  &&\phantom{aa}&[name=lbu]1,1,1 &\phantom{aa}&&\phantom{aa}&[name=rbu]0,0,0\\
  &[name=lfu]1,1,1&&&&[name=rfu]1,1,1\\
  &&&[name=lbd]0,0,0&&&&[name=rbd]1,1,1\\
  &[name=lfd]0,0,0&&&&[name=rfd]0,0,0\\
  \ncline[]{lbu}{rbu}
  \ncline[doubleline=true]{lfu}{rfu}
  \ncline[doubleline=true]{lfu}{lbu}
  \ncline[]{rfu}{rbu}
  \ncline[]{lfu}{lfd}
  \ncline[]{rfu}{rfd}
  \ncline[]{lfd}{rfd}
  \ncline[]{rbu}{rbd}
  \ncline[]{rfd}{rbd}
  \ncline[linestyle=dashed,dash=3pt 2pt]{lfd}{lbd}
  \ncline[linestyle=dashed,dash=3pt 2pt]{lbu}{lbd}
  \ncline[linestyle=dashed,dash=3pt 2pt]{lbd}{rbd}
\endpsmatrix\]

\section[Equilibrium for multi Strategic Games]{Discrete and Static Compromising Equilibrium for multi Strategic Games}\label{sect:dsceq-msg}

This section defines multi strategic games and their non deterministic strategy profiles. Then, it defines a notion of preference among sequences of sets of outcomes, which yields a notion of (global) nd equilibrium. An embedding of these multi strategic games into the ndbr multi strategic games shows that multi strategic games always have an nd equilibrium. Since sequential graph games can be embedded into multi strategic games, this also provides sequential graph games with a notion of nd equilibrium that enjoys guarantee of existence. More subtle notions of equilibrium could be defined, which would yield stronger results. However, the notion that is defined here intends to be a simple one. In addition, the section discusses a relevant embedding of strategic games into multi strategic games.

multi strategic games have a graph-like structure. At each node of the game, all agents play a strategic game to choose the outcome as well as the move to the next node. Another strategic game corresponds to this next node. So, a play in a multi strategic game is an infinite sequence of local plays in strategic games followed by outcomes. The notion of multi strategic game is defined below.

\begin{defn}[multi strategic game]
A multi strategic game is a pair $\langle S, (P^\node)^{\node\in\V}\rangle$ that complies with the following.
\begin{itemize}
\item $S=\bigotimes_{\node\in\V}\bigotimes_{a\in\Agent}S^\node_a$, where $\V$ is a non-empty set of indices, $\Agent$ is a non-empty set of agents, and $S^\node_a$ is a non-empty set of strategies.
\item  $P^\node$ is of type $S^\node\to OC\times\V$, where $\Oc$ is a non-empty set of outcomes.
\end{itemize}
The agents' (pure) strategies are the elements of $S_a=\bigotimes_{\node\in\V}S^\node_a$ and the (pure) strategy profiles are the elements of $S$.
\end{defn}

The following example depicts a multi strategic game that involves two agents, say $vertical$ and $horizontal$. At each node $vertical$ chooses the row and $horizontal$ chooses the column. The game involves natural-valued payoff functions. The first figure corresponds to agent $vertical$ and the second figure to agent $horizontal$. At a node, if a payoff function is enclosed in a small box with an arrow pointing from the bow to another node, it means that the corresponding strategy profile leads to this other node. If a payoff function is not enclosed in a small box, it means that the corresponding strategy profile leads to the same node. For instance below, If the play start at the top-left node and if the agents choose the top-left profile, then both agents get payoff $2$ and the same strategic game is played again. Whereas if the agents choose the top-right profile, $vertical$ gets payoff $2$ and $horizontal$ gets payoff $1$, and the agents have to play in the top-right node.

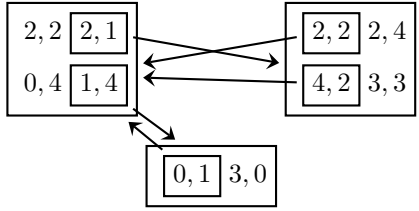
\[\psmatrix
  &[name=n1]\psframebox{
  \psmatrix
  &[name=c1] 2,2 &[name=c2] \psframebox{2,1}\\
  &[name=c3] 0,4  &[name=c4] \psframebox{1,4}
    \endpsmatrix}
  & 
 &[name=n2]\psframebox{
  \psmatrix
  &[name=c5] \psframebox{2,2}  &[name=c6] 2,4\\
  &[name=c7] \psframebox{4,2} &[name=c8] 3,3
    \endpsmatrix}\\\\
 & & [name=n3]\psframebox{
  \psmatrix
  &[name=c9] \psframebox{0,1}  &[name=c10] 3,0
    \endpsmatrix}
 \ncline{->}{c2}{n2}
 \ncline{->}{c4}{n3}
 \ncline{->}{c7}{n1}
 \ncline{->}{c5}{n1}
 \ncline{->}{c9}{n1}
 \endpsmatrix\]

Starting from every node, a strategy profile induces infinite sequences of outcomes. It is therefore possible to define local/global equilibria for multi strategic games in the same way that they are defined for sequential graph games. Then, it is possible to embed sequential graph games into multi strategic games in a way that preserves and reflects local/global equilibria. (The embedding consists in seeing a node of a sequential graph game as a 1-agent strategic game.) 

It is possible to embed strategic games into multi strategic games such that Nash equilibria correspond to local/global equilibria. (The embedding consists in seeing a strategic game as a one-node multi strategic game looping on itself).

However, since not all strategic games have a Nash equilibrium, not all multi strategic games have a local/global equilibrium. That is why non determinism comes into play.

\begin{defn}[non deterministic strategies and profiles]
Let $\langle S, (P^\node)^{\node\in\V}\rangle$ be a multi strategic game. Let $\Sigma^\node_a\eqdef\powerset{S^\node_a}-\{\emptyset\}$ be the set of local nd strategies of agent $a$ at node $\node$. Let $\Sigma_a\eqdef\bigotimes_{\node\in\V}\Sigma^\node_a$ be the set of nd strategies of agent $a$ (accounting for choices at all nodes). Let $\Sigma\eqdef\bigotimes_{a\in\Agent}\Sigma_a$ be the set of nd strategy profiles. Also, let $\Sigma^\node\eqdef\bigotimes_{a\in\Agent}\Sigma^\node_a$ be the set of local nd strategy profiles at node $\node$. For $\sigma$ in $\Sigma$, the objects $\sigma_a$, $\sigma^n$, and $\sigma_a^\node$ correspond to the projections of $\sigma$ on $\Sigma_a$, $\Sigma^n$, and $\Sigma_a^\node$.
\end{defn}

Consider an nd strategy profile. At a given node, the outcome and the next node that are prescribed by the profile may be undetermined, because of non determinism. However, the outcome is element of a determined set of outcomes and the next node is element of a determined set of nodes. The same phenomenon arises at each possible next nodes. Therefore, any path starting from a node and non deterministically following the prescription of the nd strategy profile yields a sequence whose $k$th element is element of a determined set of outcomes. These sets are defined below.

\begin{defn}[Induced sequence of sets of outcomes]
Let $g=\langle S, (P^\node)^{\node\in\V}\rangle$ be a multi strategic game and let $\sigma$ be in $\Sigma$. The induced sequence $seq(\sigma,\node)$ is an infinite sequence of non-empty subsets of outcomes.
\[seq(\sigma,\node)\quad\eqdef\quad fst\circ P^\node(\sigma^\node)\cdot seq(\sigma,snd\circ P^\node(\sigma^\node))\]
Where $f(Z)=\cup_{x\in Z}f(x)$ for any $f:\,X\to Y$ and $Z\subseteq X$, and the projection operations are defined such that $fst((x,y))=x$ and $snd((x,y))=y$. 
\end{defn}

Inclusion of nd strategy profiles implies component-wise inclusion of their sequences of sets of outcomes, as stated below.

\begin{lem}\label{lem:seq-incl}
Let $g=\langle S, (P^\node)^{\node\in\V}\rangle$ be a multi strategic game. 
\[\sigma\subseteq\sigma'\quad\Rightarrow\quad\forall\node\in\V,\forall k\in\mathbb{N},\quad seq(\sigma,\node)(k)\subseteq seq(\sigma',\node)(k)\]
\end{lem}

For every agent, a preference binary relation over outcomes can be extended to a preference over sequences of sets of outcomes, as defined below. This extended preference amounts to component-wise preference for all sets in the sequence. However, there are other "natural" ways of extending preference over outcomes to preference over sequences of sets of outcomes (using limit sets or lexicographic ordering, for instance). So, what follows is only an example.

\begin{defn}[Preference over sequences]
Let $\prec$ be a binary relation over a set $E$. It can be extended to non-empty subsets of $E$ as follows.
\[X\prec^{set}Y\quad\eqdef\quad\forall x\in X,\forall y\in Y,\quad x\prec y\]
This can be extended to sequences of non-empty subsets of $E$. Below, $\alpha$ and $\beta$ are sequences of non-empty subsets of $E$.
\[\alpha\prec^{fct}\beta\quad\eqdef\quad\forall k\in\mathbb{N},\quad\alpha(k)\prec^{set}\beta(k)\] 
\end{defn}

The definition of $\prec^{set}$ implies the following result.

\begin{lem}\label{lem:subset-prec-set}
$\emptyset\neq X'\subseteq X\quad\wedge\quad\emptyset\neq Y'\subseteq Y\quad\wedge\quad X\prec^{set}Y\quad\Rightarrow\quad X'\prec^{set}Y'$
\end{lem}

Next lemma states that the preference extension $\prec^{fct}$ preserves an ordering property.

\begin{lem}\label{lem:spo-fct}
If $\prec$ is a strict partial order then $\prec^{fct}$ is also a strict partial order.
\end{lem}

The following result shows a preservation property involving $\prec^{fct}$ and strategy inclusion.

\begin{lem}\label{lem:incl-pref-fct}
Let $g=\langle S, (P^\node)^{\node\in\V}\rangle$ be a sequential graph game. Assume that $\gamma\subseteq\delta$ are both in $\Sigma_{-a}$. Then the following holds.
\[\aPref{fct}{a}{seq(\delta;\sigma_a,\node)}{seq(\delta;\sigma_a',\node)}\quad\Rightarrow\quad\aPref{fct}{a}{seq(\gamma;\sigma_a,\node)}{seq(\gamma;\sigma_a',\node)}\]
\end{lem}

\begin{proof}
By assumption and definition, $\aPref{set}{a}{seq(\delta;\sigma_a,\node)(k)}{seq(\delta;\sigma_a',\node)(k)}$ for all $k$ in $\mathbb{N}$. Since $\gamma\subseteq\delta$, lemma~\ref{lem:seq-incl} implies that $seq(\gamma;\sigma_a,\node)(k)\subseteq seq(\delta;\sigma_a,\node)(k)$ and $seq(\gamma;\sigma'_a,\node)(k)\subseteq seq(\delta;\sigma'_a,\node)(k)$. Since these sets are all non-empty and by lemma~\ref{lem:subset-prec-set}, $\aPref{set}{a}{seq(\gamma;\sigma_a,\node)(k)}{seq(\gamma;\sigma_a',\node)(k)}$ for all $k$. 
\end{proof}

Below, multi strategic game are embedded into ndbr multi strategic games. The corresponding ndbr multi strategic games always has an ndbr equilibrium, which is interpreted as multi strategic games always having an nd equilibrium. However, the embedding is not the only relevant one. It is intended to be a simple one. More subtle embeddings can yield to stronger results of nd equilibrium existence.

\begin{lem}\label{lem:ndbr multi Nash-eq-ex}
Let $g=\langle S, (P^\node)^{\node\in\V}\rangle$ be a multi strategic game. For each agent $a$, assume that his preference over outcomes $\aPrefFct{}{\agent}$ are strict partial orders, \textit{i.e.} irreflexive and transitive. For each agent $a$ and each $\gamma$ in $\Sigma_{-a}$, the following defines an element of $\Sigma_a$.
\[BR_a(\gamma)\quad\eqdef\quad\bigotimes_{i\in I}BR_a^\node(\gamma)\]
Where
\[\]
\begin{eqnarray*}
BR_a^\node(\gamma)\quad\eqdef\quad\{s^\node\,\mid\,s\in S_a\quad\wedge\quad\forall s'\in S_a,\quad\neg(\aPref{fct}{a}{seq(\gamma;s,\node)}{seq(\gamma;s',\node)})\} 
\end{eqnarray*}
The object $\langle (S^\node_a)^{\node\in\V}_{a\in\Agent},(BR_a)_{a\in\Agent}\rangle$ is an ndbr multi strategic game, and it has an ndbr equilibrium. 
\end{lem}

\begin{proof}
First prove that $BR^\node_a(\gamma)$ is non-empty: the set of the $seq(\gamma;s,\node)$, when $s$ ranges over $S_a$, is finite and non-empty. So at least one of the $seq(\gamma;s,\node)$ is maximal with respect to $\aPrefFct{fct}{a}$ which is a strict partial order by lemma~\ref{lem:spo-fct}. Second, assume that $\gamma\subseteq\delta$ for $\gamma$ and $\delta$ in $\Sigma_{-a}$ and prove $BR^\node_a(\gamma)\subseteq BR^\node_a(\delta)$: let $c$ be in $BR^\node_a(\gamma)$. By definition, $c=s^\node$ for some strategy $s$ in $S_a$ such that $\neg(\aPref{fct}{a}{seq(\gamma;s,\node)}{seq(\gamma;s',\node)})$ for all $s'$ in $S_a$. So $\neg(\aPref{fct}{a}{seq(\delta;s,\node)}{seq(\delta;s',\node)})$ for all $s'$ in $S_a$, by contraposition of lemma~\ref{lem:incl-pref-fct}. So $c$ belongs to $BR^\node_a(\delta)$. Therefore, the ndbr multi strategic game has an ndbr equilibrium by lemma~\ref{lem:ndne}. 
\end{proof}

In the lemma above, the definition of $BR$ says that at a given node and in a given context (other agents have chosen their nd strategies), an agent dismisses any of its options that induces a sequence worse than a sequence induced by some other option. Since the result above is constructive, it provides an algorithm for finding an nd equilibrium. An example is given below. The game involves two agents, namely $vertical$ and $horizontal$. Agent $vertical$ chooses the rows and is rewarded with the first figures given by the payoff functions. 

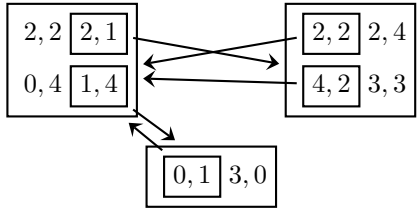
\[\psmatrix
  &[name=n1]\psframebox{
  \psmatrix
  &[name=c1] 2,2 &[name=c2] \psframebox{2,1}\\
  &[name=c3] 0,4  &[name=c4] \psframebox{1,4}
    \endpsmatrix}
  & 
 &[name=n2]\psframebox{
  \psmatrix
  &[name=c5] \psframebox{2,2}  &[name=c6] 2,4\\
  &[name=c7] \psframebox{4,2} &[name=c8] 3,3
    \endpsmatrix}\\\\
 & & [name=n3]\psframebox{
  \psmatrix
  &[name=c9] \psframebox{0,1}  &[name=c10] 3,0
    \endpsmatrix}
 \ncline{->}{c2}{n2}
 \ncline{->}{c4}{n3}
 \ncline{->}{c7}{n1}
 \ncline{->}{c5}{n1}
 \ncline{->}{c9}{n1}
 \endpsmatrix\]

In the beginning, all agents consider all of their options. At the bottom node, only agent $horizontal$ has an actual decision to take. If he chooses right, he gets an infinite sequence of $0$. ($vertical$ gets an infinite sequence of $3$, but $horizontal$ does not take it into account.) If he chooses left, he gets an infinite sequence of (non-zero) positive numbers whatever $vertical$'s strategy may be, which is better than $0$ at any stage of the sequence. So $horizontal$ dismisses his right strategy at the bottom node, as depicted below.

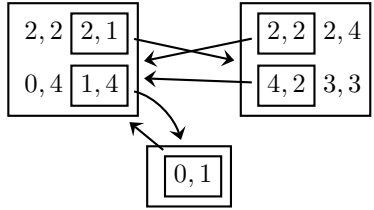
\[\psmatrix
  &[name=n1]\psframebox{
  \psmatrix
  &[name=c1] 2,2 &[name=c2] \psframebox{2,1}\\
  &[name=c3] 0,4  &[name=c4] \psframebox{1,4}
    \endpsmatrix}
  & 
 &[name=n2]\psframebox{
  \psmatrix
  &[name=c5] \psframebox{2,2}  &[name=c6] 2,4\\
  &[name=c7] \psframebox{4,2} &[name=c8] 3,3
    \endpsmatrix}\\\\
 & & [name=n3]\psframebox{
  \psmatrix
  &[name=c9] \psframebox{0,1}
    \endpsmatrix}
 \ncline{->}{c2}{n2}
 \ncarc[arcangleA=30,arcangleB=30]{->}{c4}{n3}
 \ncline{->}{c7}{n1}
 \ncline{->}{c5}{n1}
 \ncline{->}{c9}{n1}
 \endpsmatrix\]

Now agent $vertical$ considers the top-left node. If he chooses his bottom strategy, the induced sequence involves only $0$ and $1$. If he choose his top strategy, the induced sequence involves only numbers that are equal to or greater than $2$, which is better. So $vertical$ dismisses his top strategy at the top-left node node, as depicted below.

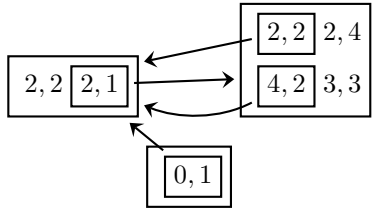
\[\psmatrix
  &[name=n1]\psframebox{
  \psmatrix
  &[name=c1] 2,2 &[name=c2] \psframebox{2,1}
    \endpsmatrix}
  & 
 &[name=n2]\psframebox{
  \psmatrix
  &[name=c5] \psframebox{2,2}  &[name=c6] 2,4\\
  &[name=c7] \psframebox{4,2} &[name=c8] 3,3
    \endpsmatrix}\\\\
 & & [name=n3]\psframebox{
  \psmatrix
  &[name=c9] \psframebox{0,1}
    \endpsmatrix}
 \ncline{->}{c2}{n2}
 \ncarc[arcangleA=30,arcangleB=20]{->}{c7}{n1}
 \ncline{->}{c5}{n1}
 \ncline{->}{c9}{n1}
 \endpsmatrix\]

Now agent $horizontal$ considers the top-right node. If he chooses his left strategy, the induced sequence involves only $1$ and $2$. If he choose his right strategy, the induced sequence involves only $3$ and $4$, which is better. So agent $horizontal$ dismisses his left strategy at the top-right node, as depicted below.

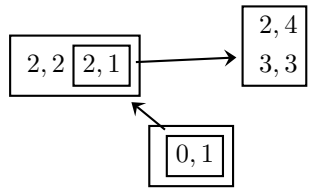
\[\psmatrix
  &[name=n1]\psframebox{
  \psmatrix
  &[name=c1] 2,2 &[name=c2] \psframebox{2,1}
    \endpsmatrix}
  & 
 &[name=n2]\psframebox{
  \psmatrix
  &[name=c6] 2,4\\
  &[name=c8] 3,3
    \endpsmatrix}\\\\
 & & [name=n3]\psframebox{
  \psmatrix
  &[name=c9] \psframebox{0,1}
    \endpsmatrix}
 \ncline{->}{c2}{n2}
 \ncline{->}{c9}{n1}
 \endpsmatrix\]

Eventually, agent $vertical$ dismisses one of his strategy at the top-right node, which yields the global nd equilibrium below. Said otherwise, for each agent, fore each node, the agent cannot get better sequence by changing his strategy. 

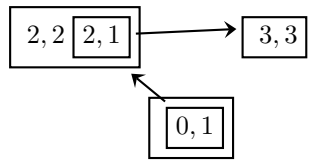
\[\psmatrix
  &[name=n1]\psframebox{
  \psmatrix
  &[name=c1] 2,2 &[name=c2] \psframebox{2,1}
    \endpsmatrix}
  & 
 &[name=n2]\psframebox{
  \psmatrix
  &[name=c8] 3,3
    \endpsmatrix}\\\\
 & & [name=n3]\psframebox{
  \psmatrix
  &[name=c9] \psframebox{0,1}
    \endpsmatrix}
 \ncline{->}{c2}{n2}
 \ncline{->}{c9}{n1}
 \endpsmatrix\]

Since sequential graph games can be embedded into multi strategic games, lemma~\ref{lem:ndbr Nash-eq-ex} also provides a notion of global nd equilibrium for sequential graph games, with the guarantee of equilibrium existence. However, it is also possible to define more subtle notions of equilibrium with this guarantee. 

\section{Conclusion}

This paper introduces the notion of abstract strategic game, which is a natural and minimalist abstraction of traditional strategic games with real-valued payoff functions. It also defines the notion of multi strategic game which is a generalisation of both abstract strategic games and sequential graph games. multi strategic games can therefore model decision-making problems that are modelled by either strategic games or sequential graph/tree games. Since these new games can express both sequential and simultaneous decision-making within the same game, they can also model more complex decision-making problems. The paper also defines non deterministic best response multi strategic games. While somewhat more abstract, they are structurally similar to multi strategic games: Cartesian product and graph-like structure. Via a pre-fixed point result that is also proved in the paper, existence of ndbr equilibrium is guaranteed for ndbr multi strategic games (under some sufficient condition). Instantiating this result with (more) concrete best response functions provides different notions of non deterministic equilibrium for multi strategic games. A few examples show the effectiveness of the approach, in terms of numerical result as well as algorithmic complexity (polynomial and low). This approach is discrete and static, so it lies between Nash's probabilistic approach and the CP an BR approach.

\bibliographystyle{plain}
\bibliography{discrete_non_det_report}

\end{document}